\documentclass[11pt]{article}

\usepackage{amssymb}
\usepackage{amsmath}
\usepackage{lineno}
\usepackage{graphicx}
\usepackage{bm}
\newcommand{\mathsym}[1]{{}}

\usepackage{graphicx}
\usepackage{subcaption} 
\usepackage{rotating}
\usepackage{amsmath}
\usepackage{color}
\usepackage{cancel}

\usepackage{hyperref}
\hypersetup{
  colorlinks   = true, 	
  urlcolor     = blue, 	
  linkcolor    = blue, 	
  citecolor   = magenta 	
}
\usepackage{cite}

\newcommand{\bra}{\begin{array}}
\newcommand{\era}{\end{array}}
\newcommand{\beq}{\begin{equation}}
\newcommand{\eeq}{\end{equation}}
\newcommand{\beqar}{\begin{eqnarray}}
\newcommand{\eeqar}{\end{eqnarray}}

\newcommand{\be}{\begin{equation}}
\newcommand{\ee}{\end{equation}}
\newcommand{\bea}{\begin{eqnarray}}
\newcommand{\eea}{\end{eqnarray}}
\newcommand{\bd}{\begin{displaymath}}
\newcommand{\ed}{\end{displaymath}}

\title{
	\vskip-1.5cm
  Analytical Solutions to Asymmetric Two-Photon Rabi Model }

\author{
	M. Baradaran$^1$\footnote{marzieh.baradaran@uhk.cz, ORCID: \href{http://orcid.org/0000-0002-8455-9973}{0000-0002-8455-9973}}, 
	L.M. Nieto$^2$\footnote{luismiguel.nieto.calzada@uva.es, ORCID: \href{http://orcid.org/0000-0002-2849-2647}{0000-0002-2849-2647}}, 
	and 
	S. Zarrinkamar$^{2,3}$\footnote {saber.zarrinkamar@uva.es, ORCID: \href{http://orcid.org/0000-0001-9128-4624}{0000-0001-9128-4624}}
	\\  [0.7ex]
	\small
	$^1$\,Department of Physics, Faculty of Science, University of Hradec Kr\'alov\'e,\\
	\small
	Rokitansk\'eho 62, 500 03 Hradec Kr\'alov\'e, Czechia
	\\ [0.1ex]
	\small
	$^2$\,Departamento de F\'{\i}sica Te\'{o}rica, At\'{o}mica y \'{O}ptica, and Laboratory for Disruptive \\ 
		\small
		Interdisciplinary Science (LaDIS), Universidad de Valladolid, 47011 Valladolid, Spain
	\\ [0.1ex]
	\small
	$^3$\,Departament of Basic Sciences, Garmsar Branch,
	Islamic Azad University, Garmsar, Iran
}

\begin{document}
	
	\maketitle

\begin{abstract}

Within the Segal-Bargmann representation, a generalized Rabi model is considered that includes both two-photon and asymmetric terms. It is shown that, through a suitable transformation, nearly exact solutions can be obtained using the Bethe ansatz approach. Applying this approach to the meromorphic structure of the resulting differential equation, solutions in exact analytical form of the fourth-order problem are presented for both an arbitrary state and for the restriction between the parameters.

\end{abstract}


\section{Introduction}
After several decades, Rabi's ingenious proposal for the field-atom interaction \cite{Rabi} remains one of the most attractive models in quantum optics and related fields, including quantum communication, laser physics, and quantum technologies. This could be due to several reasons, including the simple yet rich structure of the proposed Hamiltonian, as well as several challenges with other parallel approaches. Recently, there has been renewed interest in the so-called Segal-Bargmann representation \cite{Bargman61,Segal63}  of the Rabi Hamiltonian, which could be particularly interesting because of its expression as a differential equation and the more familiar form of the solutions for much of the community working on these topics. Examples of pioneering work in this area are those of Koc   \cite{Koc 2002} and especially the impactful work of Braak \cite{Braak 2011 PRL}, whose ideas were soon applied to other versions and generalizations of the model,  including two-photon, two-mode, two-qubit, asymmetric, anisotropic, driven, and other forms \cite{Travenec 2012, Xie 2013, Zhong 2013, Li 2015, Maciejewski 2017, Maciejewski 2019, Xie Asymmetric two-photon, Arxiv 2024, Braak 2024}. In particular, the two-photon Rabi and asymmetric Rabi models \cite{Xie 2013, Xie Asymmetric two-photon}, due to their remarkable applications in ion traps, laser fields, qubit circuits and others \cite{Felicetti 2015, Nature Photonics 2016, Nature Physics 2017}, have been the subject of very recent analytical and experimental studies \cite{Chen 2018, Polaron PRA 2019, Liu 2022, Quantum electronics 2024, Chan 2020 JPA, Rico Collapse PRA 2020, Lo 2022}.

The problems with the Segal-Bargmann representation are numerous. The first is that the associated equation becomes quite complex to handle when we consider the case of $n$ photons, since we must deal with a differential equation of order $2n$. Specifically, in the case of two photons, we encounter a fourth-order differential equation that has not been thoroughly analyzed. This might seem strange, but it is precisely that, since we normally (though not always) deal with a second-order differential equation problem, of a linear nature, ranging from Newton's classical equation to the wave equations of quantum mechanics, whether relativistic or non-relativistic.
While some attempts have been made to address these problems, including those arising within the Generalized Uncertainty Principle or the Minimum Length formalism \cite{Vagenas PRL 2008, Vagenas 2009, Review 2015}, the field remains quite open to investigation from an analytical perspective, even with seemingly simple cases such as $n=2$. Undoubtedly, the case for higher numbers, i.e., $n\geq 3$, analyzed in Braak's recent article \cite{Braak 2024}, presents a real challenge.

Quite similar problems arise for the multi-qubit and multimode cases, although the two-mode case is addressed in an interesting way by Zhang \cite{Zhang 2013}. Even in some generalizations of the model with the two-photon term, the resulting fourth-order differential equation and its second-order approximation are not so easy to solve.
The problem is also not straightforward in the case of generalized Rabi models for two photons, modified with asymmetric or anisotropic terms, and it is necessary to go beyond the common Lie algebraic approach \cite{Turbiner88, Artemio94, Turbiner} and Heun forms \cite{Ronveaux, Ishkhanyan 2015 EPJD, Ishkhanyan 2018}. This is achieved through the interesting pioneering work of Zhang and collaborators \cite{Zhang2012, Zhang 2013, Zhang 2013 AOP, Zhang AOP, Zhang 2017, Moroz 2018}  for various generalizations of Rabi's basic model.

In Section~\ref{22}, we consider a generalization of Rabi's model that includes biphotonic and asymmetric terms and show that the problem, when transformed to a Segal-Bargmann space, admits analytical solutions. 
After some calculations, we present the solutions for an arbitrary state and conclude the work with some final remarks.

\section {The model and its analysis}\label{22}

The Hamiltonian that will be studied is the following
\be
H_0=\Delta \sigma_z+\epsilon \sigma_x+ \omega a^\dagger a+\lambda \sigma_x \left({a^\dagger}^2+a^2\right),
\ee
where $a$ and $a^\dagger$  denote, respectively, the annihilation and creation operators, $\omega$ represents the frequency of the single bosonic mode, $\lambda$ denote the interaction squeezing, $\sigma_x$ and $\sigma_z$ are Pauli matrices for the two-level system with level splitting $2\Delta$,  and the term $\epsilon \sigma_x$ represents the spontaneous transition of the two-level system. Obviously, the model is a generalization of both two-photon Rabi model and asymmetric Rabi models. We will work in the Segal-Bargmann space of holomorphic functions $f(z)$, $z\in\mathbb{C}$, where the creation and annihilation operators are represented by the differential operators  $a^\dagger\rightarrow z$ and $a\rightarrow \frac{d}{dz}$.
On the time-independent Schrödinger equation $H_0 |\psi\rangle =E |\psi\rangle $ we perform the unitary transformation  \cite {Arxiv 2024}
\be
|\psi\rangle =\left( \begin{array}{c}\psi_1(z)\\ \psi_2(z) \end{array} \right)=U\left( \begin{array}{c}\phi_1(z) \\ \phi_2(z) \end{array} \right)=U |\phi\rangle , \quad U\equiv \frac{1}{\sqrt 2}\left(\sigma_z+\sigma_x\right),
\ee
to have $H |\phi\rangle =E |\phi\rangle $, where the new Hamiltonian is written as
\be
H=\Delta \sigma_x+\epsilon \sigma_z+ \omega a^\dagger a+\lambda \sigma_z \left({a^\dagger}^2+a^2\right).
\ee
By decoupling the components of the two-level system we arrive at \cite {Arxiv 2024}
\begin{eqnarray} \label{eqBef}
\!\!\!\!
&&\!\!\frac{d^4\phi_1(z)}{dz^4}+\left( \frac{ 2\omega+\epsilon+E}{\lambda} +\left(1-\frac{\omega^2}{\lambda^2}\right)z^2\right)\frac{d^2\phi_1(z)}{dz^2}\\   
\!\!\!\!
  && \qquad\quad \ +\left( \frac{\omega(\epsilon+E)-\omega^2}{\lambda^2} z+\frac{\omega}{\lambda}z^3\right)\frac{d\phi_1(z)}{dz}+\left(\frac{2\lambda^2+\epsilon^2-E^2-\Delta^2}{\lambda^2}+\frac{2(\epsilon-\omega)}{\lambda} z^2+z^4\right)\phi_1 (z) =0.
\nonumber	
\end{eqnarray}
We now intend to analyze this equation without any approximation.

\subsection {Bethe ansatz solutions}
\begin{figure}[htb]
	\centering
	\includegraphics[width=0.78\textwidth]{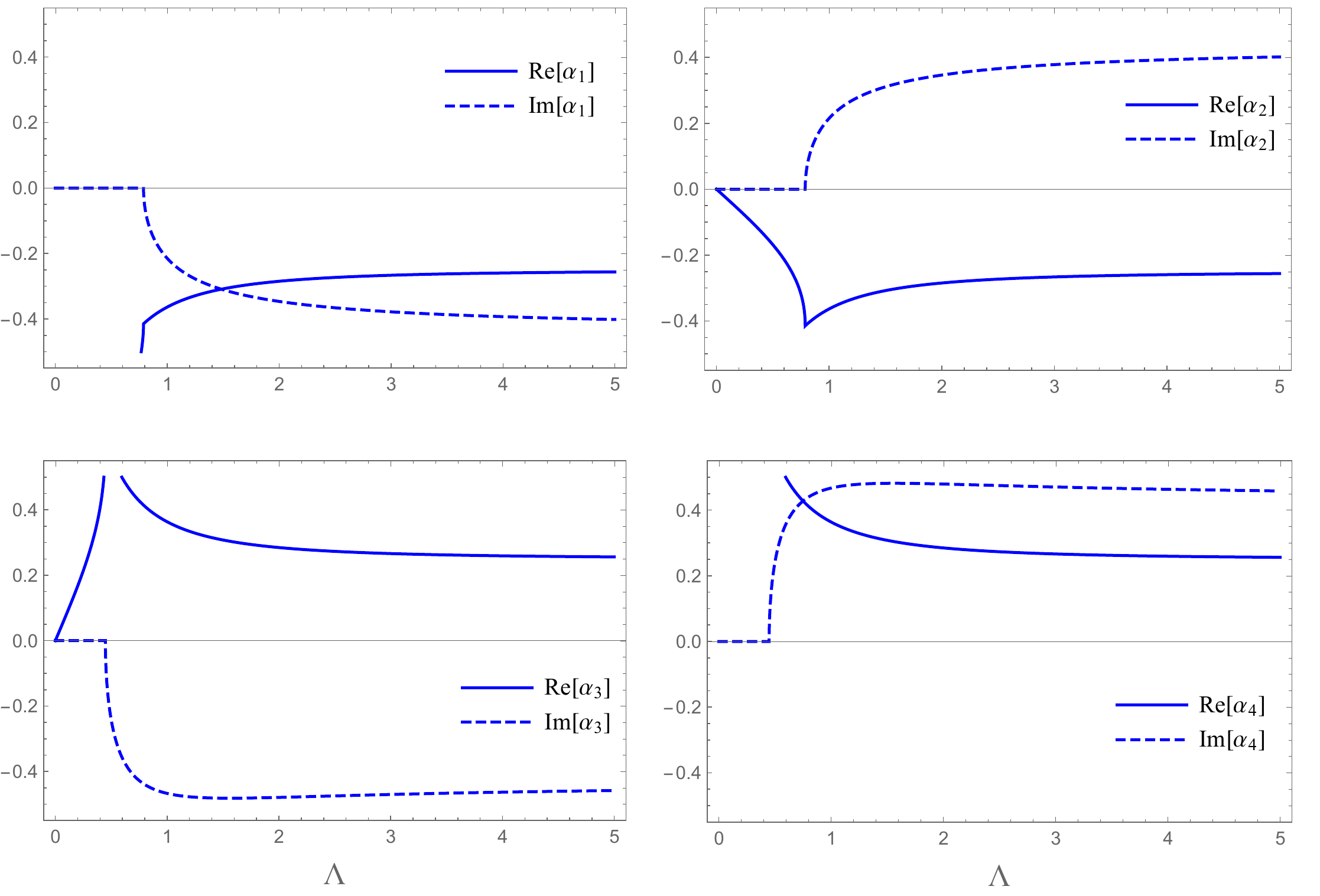}  
	\caption{\small  Real and imaginary components of $\alpha_i$, $i=1,...,4$ as a function of $\Lambda=\frac{\lambda}{\omega}$, in the range $|\alpha_i| < \frac{1}{2}$.}
	\label{figureAlpha}
\end{figure}

Using in \eqref{eqBef} the transformation $\phi_1(z)=e^ {\alpha z^2}\varphi(z)$,  $\alpha\in\mathbb{C}$, $|\alpha|< \frac{1}{2}$ (see details in Appendix~\ref{appendixA}), we get
	\begin{equation}\label{1steq}
	\frac{d^4\varphi (z)}{dz^4}+    a_1 z \frac{d^3\varphi (z) }{dz^3} +\left(   b_2 z^2+b_0  \right)\frac{d^2\varphi (z) }{dz^2}
+  \left(   p_3 z^3+  p_1 z   \right)\frac{d\varphi (z) }{dz} +\left(   q_4 z^4 +q_2 z^2+ q_0   \right)\varphi (z) =0,
	\end{equation}
where
	\begin{equation}	\begin{aligned}\label{ABV}
a_1 &=8 \alpha , \\
b_2&=24 \alpha ^2-\frac{\omega ^2}{\lambda ^2}+1 ,  \\  
b_0&=\frac{12 \alpha  \lambda +E+2 \omega +\epsilon }{\lambda } ,  \\  
p_3&= 32 \alpha ^3+\alpha  \left(4-\frac{4 \omega ^2}{\lambda ^2}\right)+\frac{\omega }{\lambda },  \\  
p_1&=  \frac{(4 \alpha  \lambda +\omega ) (12 \alpha  \lambda +E-\omega +\epsilon )}{\lambda ^2},  \\  
q_4&= \frac{-4 \alpha ^2 \omega ^2+\left(16 \alpha ^4+4 \alpha ^2+1\right) \lambda ^2+2 \alpha  \lambda  \omega }{\lambda ^2},  \\  
q_2&=  \frac{2 \lambda  \left(2 \alpha ^2 \epsilon +\alpha  (2 \alpha  (12 \alpha  \lambda +E)+\lambda )+\epsilon \right)-4 \alpha  \omega ^2+2 \omega  (\alpha  (4 \alpha  \lambda +E+\epsilon )-\lambda )}{\lambda ^2} , \\  
q_0&= \frac{2 \lambda  (\alpha  (6 \alpha  \lambda +E+2 \omega )+\lambda )+2 \alpha  \lambda  \epsilon -\Delta ^2-E^2+\epsilon ^2}{\lambda ^2} . 
	\end{aligned}	\end{equation}
To employ the Bethe ansatz method \cite{Zhang2012, Zhang 2013, Zhang 2017}, the coefficient $q_4$ in \eqref{1steq} is set to zero, and therefore,
dividing by $\omega ^2\neq0$ and denoting $\Lambda:=\frac{\lambda}{\omega}$, we can simplify further to arrive at
\begin{equation}	\label{q4=0eqLambda}
\left(16 \alpha ^4+4 \alpha ^2+1\right) \Lambda ^2-4 \alpha ^2 +2 \alpha  \Lambda=0,
\end{equation}
a quartic equation in $\alpha$,  which will have four roots $\alpha_{i}(\Lambda)$, $i=1,2,3,4$, given explicitly by
\begin{equation}	\label{alpha1234}
	\begin{aligned}
		\alpha_{1,2}(\Lambda)&=  -\frac{h(\Lambda )}{4 \sqrt{6}}\mp  \sqrt{\frac{\phantom{-}6 \sqrt{6} \Lambda  f(\Lambda )- h(\Lambda )\left(4 \left(\Lambda ^2-1\right) f(\Lambda )+f(\Lambda )^2+13 \Lambda ^4-2 \Lambda ^2+1\right)}{48\;\Lambda ^2 f(\Lambda ) h(\Lambda )}} ,\\
		\alpha_{3,4}(\Lambda)&= \phantom{-}\frac{h(\Lambda )}{4 \sqrt{6}}\mp  \sqrt{\frac{-6 \sqrt{6} \Lambda  f(\Lambda )- h(\Lambda )\left(4 \left(\Lambda ^2-1\right) f(\Lambda )+f(\Lambda )^2+13 \Lambda ^4-2 \Lambda ^2+1\right) }{48\;\Lambda ^2 f(\Lambda ) h(\Lambda )}}   ,
	\end{aligned}
\end{equation}
where the first and second subscripts of $\alpha$ correspond respectively to the upper and lower signs in the right-hand side expressions, and the functions $f(\Lambda )$ and $h(\Lambda )$ are as follows
\begin{equation}	\label{f,and,h}
	\begin{aligned}
		f(\Lambda)&= \sqrt[3]{-35 \Lambda ^6+\frac{93 \Lambda ^4}{2}+3 \Lambda ^2+\frac{3}{2} \sqrt{3} \sqrt{-\Lambda ^4 \left(144 \Lambda ^8+332 \Lambda ^6-191 \Lambda ^4-76 \Lambda ^2+20\right)}-1}  ,\\
		h(\Lambda)&=\sqrt{\frac{2 \left(-2 \Lambda ^2 f(\Lambda )+f(\Lambda )^2+2 f(\Lambda )+13 \Lambda ^4-2 \Lambda ^2+1\right)}{\Lambda ^2 f(\Lambda )}} .
	\end{aligned}
\end{equation}
Figure \ref{figureAlpha} shows the real and imaginary parts of the four roots $\alpha_i(\Lambda)$ as a function of $\Lambda=\frac{\lambda}{\omega}$, within the region $|\alpha_i| < \frac{1}{2}$, 
a restriction that is imposed by the normalization condition of the wave function.

Consequently, equation \eqref{1steq} reduces to
	\begin{equation}	 \label{2ndeq}
		\frac{d^4\varphi(z)  }{dz^4}+    a_1 z \frac{d^3\varphi(z)  }{dz^3} +\left(   b_2 z^2+b_0  \right)\frac{d^2\varphi(z) }{dz^2}
	+  \left(   p_3 z^3+  p_1 z   \right)\frac{d\varphi(z) }{dz} +\left(  q_2 z^2+ q_0   \right)\varphi (z) =0,
 	\end{equation}
where the coefficients $a,\;b, \;p$ and $q$ are given by \eqref{ABV}. Now, we propose polynomial solutions of degree $n$ in the form  \cite{Zhang2012, Zhang 2013, Zhang 2017}
\begin{equation}\label{AnsatzFunc}
	\varphi (z)  :=\varphi_n(z) =\displaystyle\prod_{i=1}^n (z-z_i),\qquad \varphi_0 \equiv 1 \quad\text{for}\quad n=0,  
\end{equation}
and we look for the values of the coefficients $q_2$ and $q_0$ in \eqref{2ndeq} such that \eqref{AnsatzFunc} is a solution of \eqref{2ndeq}. Substituting $	\varphi_n(z) $ into \eqref{2ndeq} and dividing both sides by $	\varphi_n(z) $, we obtain
\begin{equation}	
\begin{aligned}\label{BethSubs}
		-q_0=&\sum_{i=1}^n\frac{1}{z-z_i} \sum_{ p\neq \ell\neq j\neq i}^n \frac{4}{(z_i-z_p)(z_i-z_{\ell})(z_i-z_j)}  
		+ a_1  z\sum_{i=1}^n\frac{1}{z-z_i} \sum_{  \ell\neq j\neq i}^n \frac{3}{(z_i-z_{\ell})(z_i-z_j)}\\
		&+ (b_2 z^2+b_0)\sum_{i=1}^n\frac{1}{z-z_i} \sum_{   j\neq i}^n \frac{2}{(z_i-z_j)}
		+ (p_3 z^3+p_1 z)\sum_{i=1}^n\frac{1}{z-z_i} +q_2 z^2.
\end{aligned}	\end{equation}
The left-hand side of \eqref{BethSubs} is a constant, and the right-hand side is a meromorphic function with simple poles at $z=z_i$ and singularity at $z=\infty$. We then find the residues of $-q_0$ at the simple poles $z = z_i$ as
\begin{equation}	\begin{aligned}\label{ResS}
		\text{Res}{(-q_0)}_{z = z_i}=& \sum_{ p\neq \ell\neq j\neq i}^n \frac{4}{(z_i-z_p)(z_i-z_{\ell})(z_i-z_j)}  
		+ a_1  z_i \sum_{  \ell\neq j\neq i}^n \frac{3}{(z_i-z_{\ell})(z_i-z_j)}\\
		&+ (b_2 z_i^2+b_0) \sum_{   j\neq i}^n \frac{2}{(z_i-z_j)}
		+ p_3 z_i^3+p_1 z_i.
\end{aligned}	\end{equation}
Therefore, using \eqref{BethSubs} and \eqref{ResS}, we get
\begin{equation}	\begin{aligned}\label{SminusResS}
		-q_0-\sum_{i=1}^n\frac{\text{Res}{(-q_0)}_{z = z_i}}{z-z_i}=&  \;a_1  \sum_{i=1}^n\sum_{  \ell\neq j\neq i}^n \frac{3}{(z_i-z_{\ell})(z_i-z_j)}\\
		&+ \sum_{i=1}^n \left[b_2 (z+z_i)\right] \sum_{   j\neq i}^n \frac{2}{(z_i-z_j)} 
		+ \sum_{i=1}^n \left[p_3 (z z_i+ z_i^2+ z^2)+p_1\right]+q_2 z^2.
\end{aligned}	\end{equation}
Then, after a lengthy algebra and using the identities
\begin{equation}\label {Identity}
		\sum_{i=1}^n \sum_{j\neq i}^n \frac{1}{z_i-z_j}=0, \qquad\quad  \sum_{i=1}^n \sum_{j\neq i}^n \frac{z_i}{z_i-z_j}=\frac{1}{2}n(n-1),\qquad\quad
			 \sum_{i=1}^n \sum_{\ell\neq j\neq i}^n \frac{1}{(z_i-z_{\ell})(z_i-z_j)}=0     , 
\end{equation}
we can show that
\begin{equation}	\label{SminusResS2}
		-q_0= \sum_{i=1}^n\frac{\text{Res}{(-q_0)}_{z = z_i}}{z-z_i}+ \left(q_2+n p_3 \right)z^2+\left(p_3\sum_{i=1}^n z_i\right)z 
		+ p_3\sum_{i=1}^n z_i^2+n \,p_1+n(n-1)b_2.
	\end{equation}
Equating then the corresponding powers on both sides of the equation, canceling the coefficients of $z$, $z^2$ and the residues at simple poles, we obtain the following constraints
\begin{subequations}\label{BAE1}
	\begin{align}
&	q_2 =-n\,p_3	,\label{BAE1a} \\  
&	p_3 \sum_{i=1}^n z_i =0	, \label{BAE1b}\\ 
&	q_0 =-p_3 \sum_{i=1}^n z_i^2-n\,p_1-n(n-1)b_2	,\label{BAE1c}
\end{align}
\end{subequations}
where the roots $z_i$ are determined by the Bethe ansatz equations
\begin{equation}	\begin{aligned}\label{BAEqs}
&\sum_{ p\neq \ell\neq j\neq i}^n \frac{4}{(z_i-z_p)(z_i-z_{\ell})(z_i-z_j)}  
		+ a_1  z_i  \sum_{  \ell\neq j\neq i}^n \frac{3}{(z_i-z_{\ell})(z_i-z_j)}  \\
		& \qquad\qquad\qquad +(b_2 z_i^2+b_0) \sum_{   j\neq i}^n \frac{2}{z_i-z_j} + p_3 z_i^3+p_1 z_i =0 ,\qquad\qquad  i=1,2,...,n.
\end{aligned}	\end{equation}

\subsubsection {Explicit solutions for general $n$}

\begin{figure}[htb]
	\centering
	\includegraphics[width=0.3\textwidth]{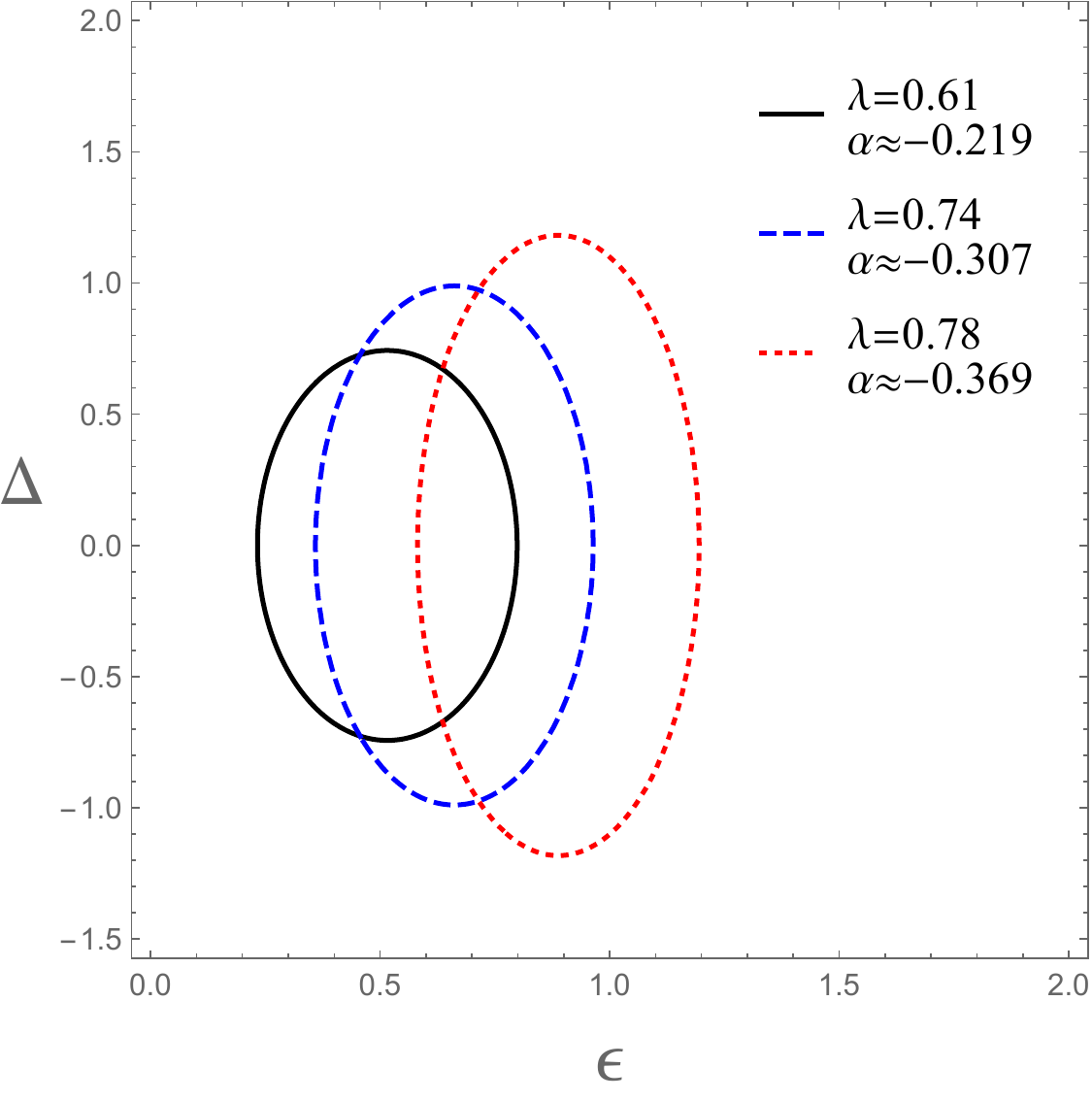}  
\caption{\small Allowed values for the parameters $\Delta$ and $\epsilon$ by \eqref{del0} when $n=0$, with $\omega=1$ and different values of $\lambda$ and $\alpha$, and taking $\alpha_2(\Lambda)$ from \eqref{alpha1234} within the region $|\alpha_2| < \frac{1}{2}$.}
	\label{En0}
\end{figure}

Using \eqref{BAE1a} together with \eqref{ABV}, we obtain the values of the energy levels $E_n$  in a closed-form as
\begin{equation}	\label{energy}
		E_n= \frac{ 4 \alpha  (n+1) \omega ^2 -2 \epsilon  \left(2 \alpha ^2 \lambda +\alpha  \omega +\lambda \right)-2 \alpha  \lambda ^2 \left(8 \alpha ^2 (2 n+3)+2 n+1\right)-\lambda  \omega  \left(8 \alpha ^2+n-2\right)}{2 \alpha  (2 \alpha  \lambda +\omega )} .	
		\end{equation}
Furthermore, from \eqref{BAE1b}--\eqref{BAE1c} and \eqref{ABV}, we obtain respectively the following two restrictions on the parameters involved:
\begin{equation}\label{constraint1}
		\left( 32 \alpha ^3+\alpha  \left(4-\frac{4 \omega ^2}{\lambda ^2}\right)+\frac{\omega }{\lambda } \right)\sum_{i=1}^n z_i =0 
	\end{equation}
and
\begin{equation}	\begin{aligned}\label{constraint2}
		\Delta ^2=&\,   n \omega\,E_n -E_n^2-\lambda ^2 \left(n-n^2-2+\frac{4 \alpha  \omega ^2-4 \left(8 \alpha ^3+\alpha \right) \lambda ^2-\lambda  \omega }{\lambda ^2}\sum_{i=1}^n z_i^2\right)  \\
		& +\epsilon ^2+12 \alpha ^2 \lambda ^2 (2 n (n+1)+1)  -n^2 \omega ^2 +2 \alpha  \lambda  (2 n+1) (E_n+2 \omega )  
	 +\epsilon  (2 \alpha  (\lambda +2 \lambda  n)+n \omega ).
\end{aligned}	\end{equation}
The roots $z_i$, $ i=1,2,...,n$, are explicitly determined from \eqref{BAEqs} and \eqref{ABV} by the set of equations 
\begin{equation}	\begin{aligned}\label{BAEqsEXPLicit}
				& \sum_{ p\neq \ell\neq j\neq i}^n \frac{4}{(z_i-z_p)(z_i-z_{\ell})(z_i-z_j)}  
		+ 8\alpha  z_i  \sum_{  \ell\neq j\neq i}^n \frac{3}{(z_i-z_{\ell})(z_i-z_j)}
		\\ 
		&\qquad\qquad +\left( (24 \alpha ^2-\frac{\omega ^2}{\lambda ^2}+1) z_i^2+\frac{12 \alpha  \lambda +E_n+2 \omega +\epsilon }{\lambda } \right) \sum_{   j\neq i}^n \frac{2}{z_i-z_j}
		\\  
		&\qquad\qquad+ \left( 32 \alpha ^3+\alpha  \left(4-\frac{4 \omega ^2}{\lambda ^2}\right)+\frac{\omega }{\lambda }\right) z_i^3+   \frac{(4 \alpha  \lambda +\omega ) (12 \alpha  \lambda +E_n-\omega +\epsilon )}{\lambda ^2}  z_i =0 .
	\end{aligned}	\end{equation}
Next, we present the explicit results for $n=0, 1$ and illustrate them graphically for the allowed parameter values. For $n=0$, from \eqref{energy} and \eqref{constraint2}, the ground state energy and the parameter constraint are 
\begin{eqnarray}\label{E0} 
&&E_0=\frac{\omega  (\lambda -\alpha  (4 \alpha  \lambda +\epsilon ))-\lambda  \left(24 \alpha ^3 \lambda +2 \alpha ^2 \epsilon +\alpha  \lambda +\epsilon \right)+2 \alpha  \omega ^2}{\alpha  (2 \alpha  \lambda +\omega )}		,   
\\
\label{del0} 
&& \Delta ^2=2 \lambda  (\alpha  (6 \alpha  \lambda +E_0+2 \omega )+\lambda )+2 \alpha  \lambda  \epsilon -E_0^2+\epsilon ^2.
\end{eqnarray}
\begin{figure}[htb]
	\centering
	\begin{subfigure}[b]{0.33\textwidth}
		\centering
		\includegraphics[width=\textwidth]{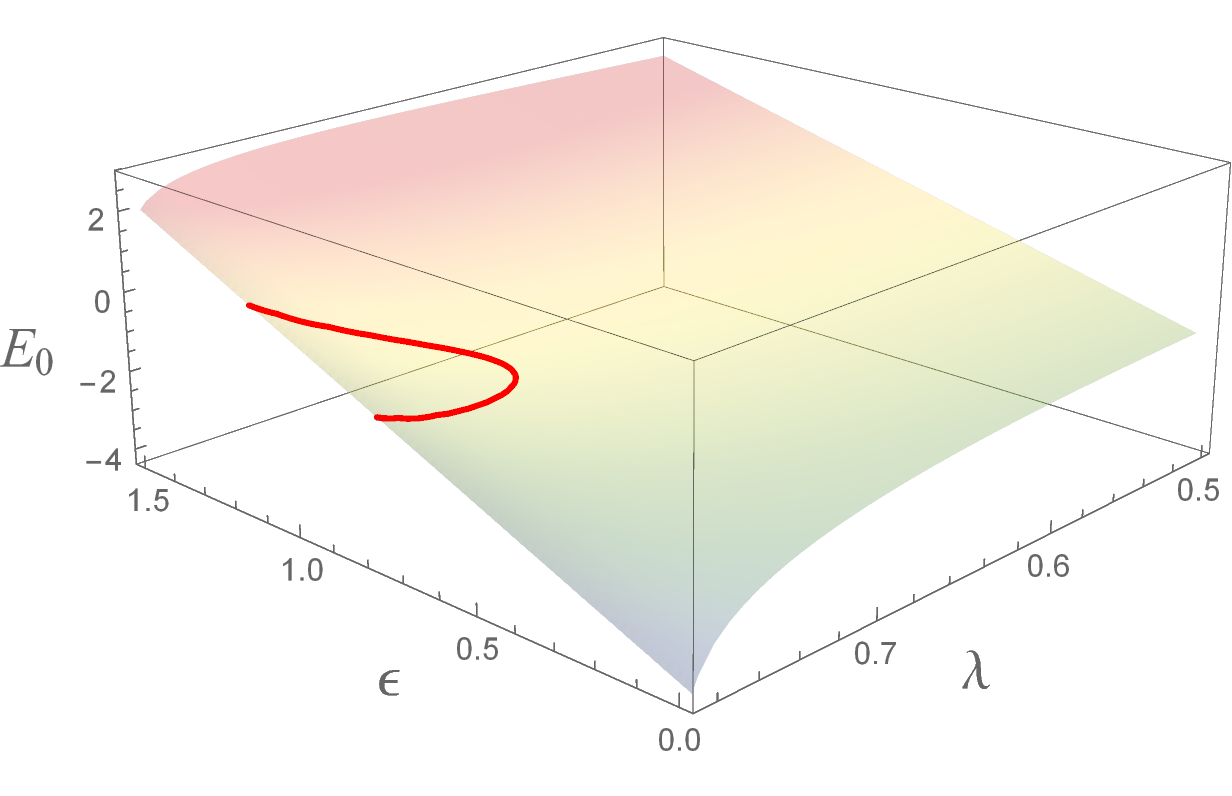}
		\caption{ $\Delta=\omega=1$.}
		\label{E0Delta1Omega1}
	\end{subfigure}
\qquad\qquad
	\begin{subfigure}[b]{0.33\textwidth}
		\centering
		\includegraphics[width=\textwidth]{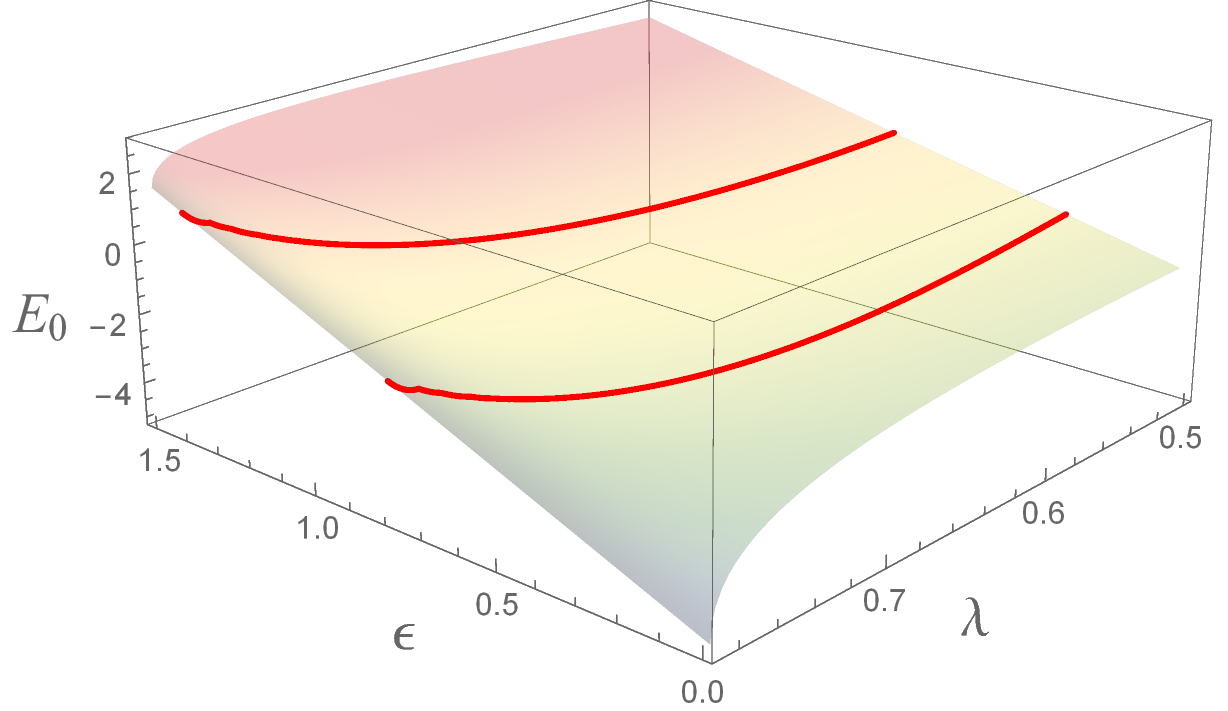}
	\caption{ $\Delta=\frac13,\;\omega=1$.}
		\label{E0Delta13rdOmega1}
	\end{subfigure}
	\caption{\small The ground state energy $E_0$ \eqref{E0}, indicated by the surfaces, and the  allowed values of the parameter \eqref{del0}, indicated by the red curves embedded in the surfaces, as functions of $\epsilon$ and $\lambda$ for fixed values of $\Delta$ and $\omega$. Here, $\alpha_2(\Lambda)$ is taken from \eqref{alpha1234} within the allowed region $|\alpha_2| < \frac{1}{2}$.}
	\label{E0Delta}
\end{figure}
Graphs of the energy \eqref{E0} and the parameter constraint \eqref{del0} as functions of $\lambda$ and $\epsilon$ are shown in Fig.~\ref{E0Delta}
 for various  values of $\Delta$ and $\omega$. 
The  constraint \eqref{del0} as a function of $\Delta$ and $\epsilon$ is illustrated in Fig.~\ref{En0}.

\begin{figure}[htb]
	\centering
	\begin{subfigure}[b]{0.37\textwidth}
		\centering
		\includegraphics[width=\textwidth]{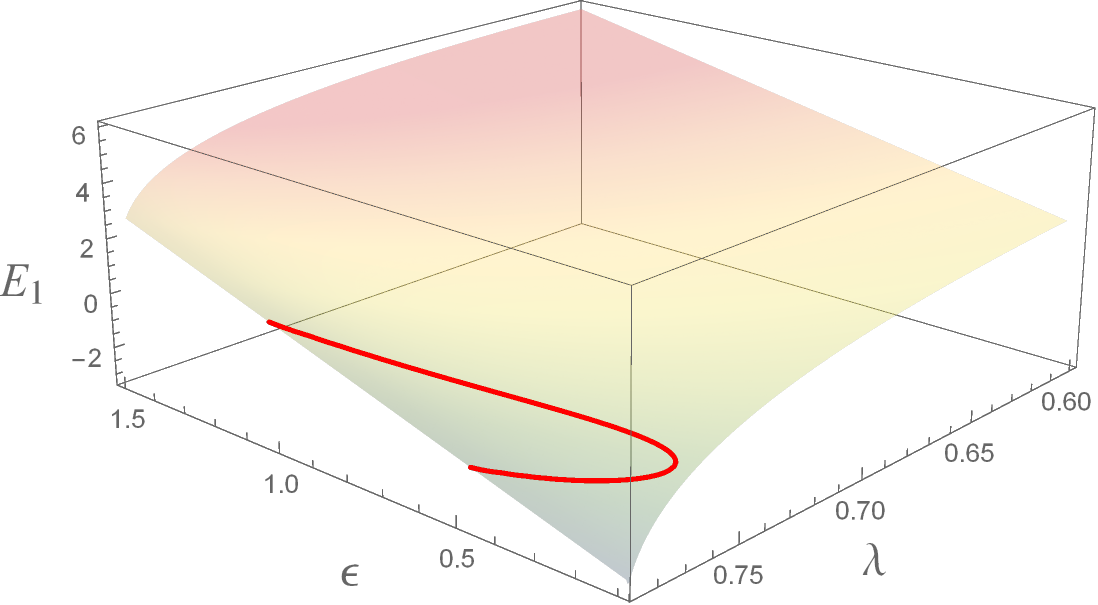}
		\caption{ $\Delta=\omega=1$ .}
		\label{E1Delta1Omega1}
	\end{subfigure}
	\qquad\qquad
	\begin{subfigure}[b]{0.34\textwidth}
		\centering
		\includegraphics[width=\textwidth]{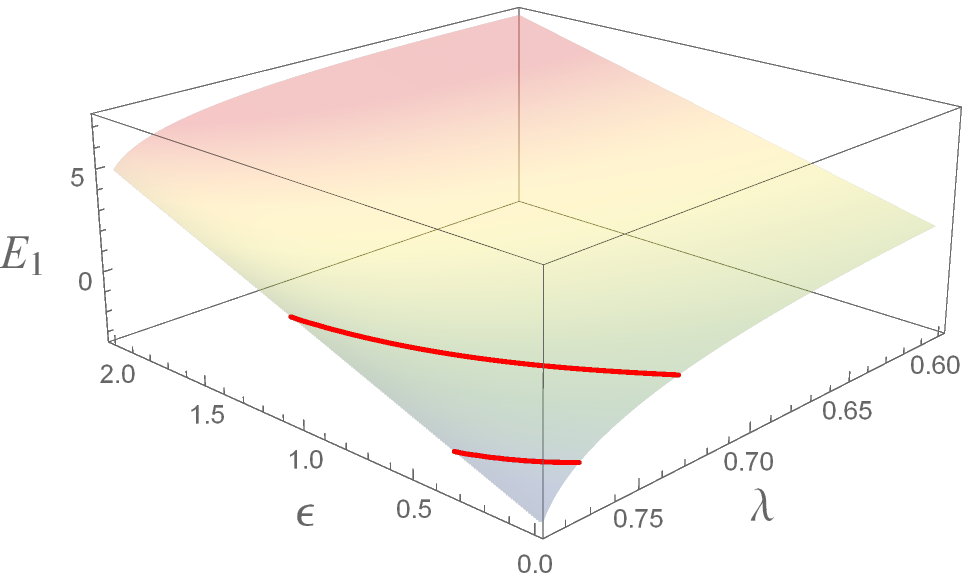}
		\caption{ $\Delta=\frac14,\;\omega=1$.}
		\label{E1Delta14thOmega1}
	\end{subfigure}
	\caption{\small 	The first excited state energy $E_1$  \eqref{E1}, indicated by the surfaces, and the  allowed values of the parameter \eqref{del1}, indicated by the red curves embedded in the surfaces, as functions of $\epsilon$ and $\lambda$ for fixed values of $\Delta$ and $\omega$. Again, $\alpha_2(\Lambda)$ is considered from \eqref{alpha1234} within the allowed region $|\alpha_2| < \frac{1}{2}$.}
	\label{E1Delta}
\end{figure}

\begin{figure}[htb]
	\centering
	\includegraphics[width=0.323\textwidth]{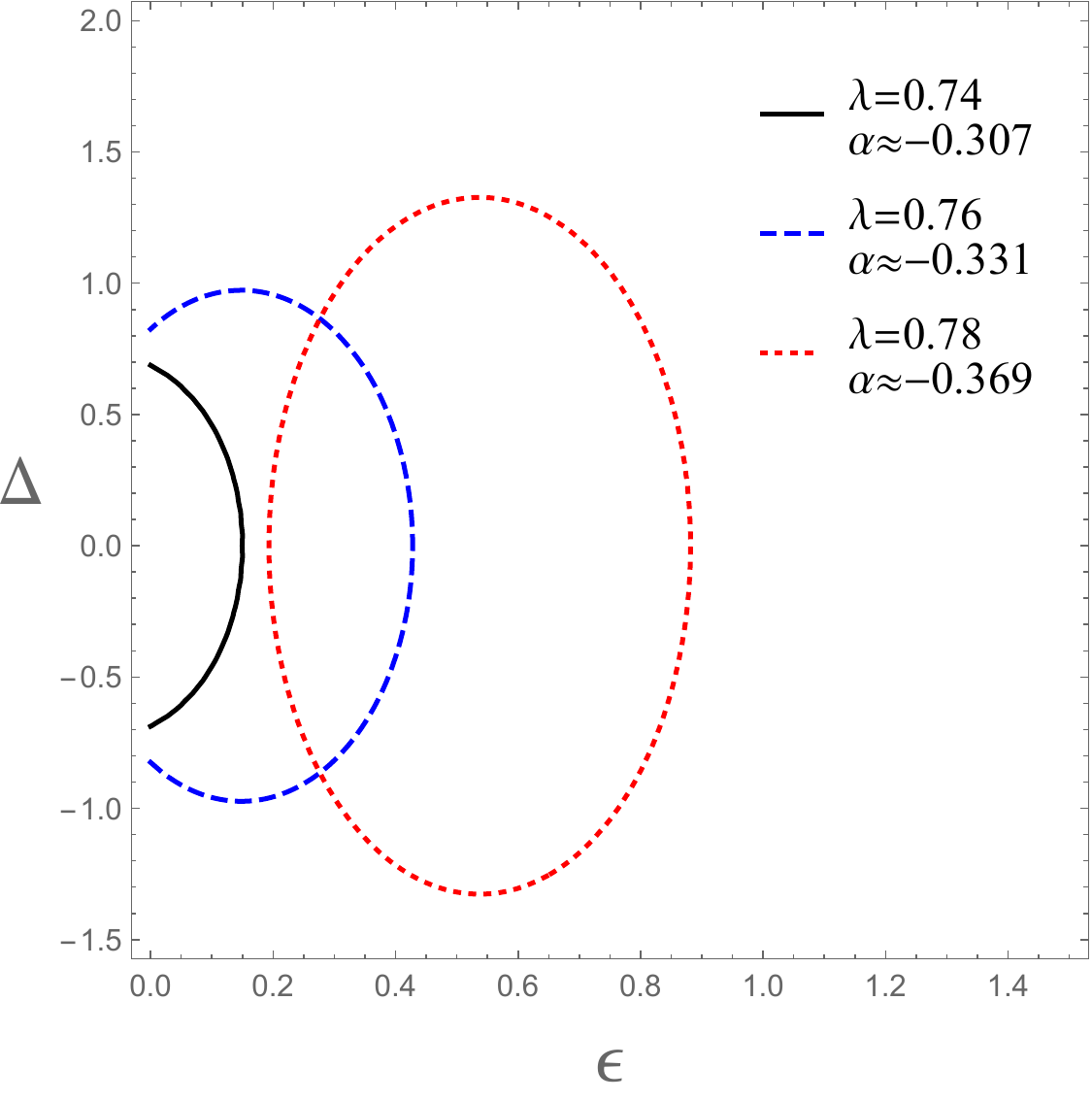}  
\caption{\small Allowed values for the parameters $\Delta$ and $\epsilon$ by \eqref{del1} when $n=1$, with $\omega=1$ and different values of $\lambda$ and $\alpha$, and taking  $\alpha_2(\Lambda)$ from \eqref{alpha1234} within the region $|\alpha_2| < \frac{1}{2}$.}
	\label{En1}
\end{figure}

Similarly, for $n=1$, from \eqref{energy}--\eqref{BAEqsEXPLicit}, we obtain the solutions for the first
excited state  in the form
	\begin{eqnarray}\label{E1}
E_1\!\!& \!\!=\!\!& \!\! \frac{\omega  (\lambda -2 \alpha  (4 \alpha  \lambda +\epsilon ))-2 \lambda  \left(\left(40 \alpha ^2+3\right) \alpha  \lambda +2 \alpha ^2 \epsilon +\epsilon \right)+8 \alpha  \omega ^2}{2 \alpha  (2 \alpha  \lambda +\omega )} , 
\\
0 \!\!& \!\!=\!\!& \!\!  \left( 32 \alpha ^3+\alpha  \left(4-\frac{4 \omega ^2}{\lambda ^2}\right)+\frac{\omega }{\lambda } \right)\, z_1   , 
\\
\nonumber
	\Delta^2 \!\!& \!\!=\!\!& \!\!  \lambda ^2 \left(60 \alpha ^2+2-\frac{4 \alpha  \omega ^2-4 \left(8 \alpha ^3+\alpha \right) \lambda ^2-\lambda  \omega }{\lambda ^2}\, z_1^2\right)+	 (6 \alpha  \lambda +\omega )E_1 
\\
&&   +12 \alpha  \lambda  \omega +\epsilon  (6 \alpha  \lambda +\omega )-E_1^2-\omega ^2+\epsilon ^2,
 \label{del1}
\end{eqnarray}
this being the Bethe ansatz equation
	\begin{equation*} 
	\left( 32 \alpha ^3+\alpha  \left(4-\frac{4 \omega ^2}{\lambda ^2}\right)+\frac{\omega }{\lambda }\right) z_1^3+   \frac{(4 \alpha  \lambda +\omega ) (12 \alpha  \lambda +E_1-\omega +\epsilon )}{\lambda ^2}  z_1 =0.
	\end{equation*}
Graphical results for the energy of the first excited state \eqref{E1} and for the parameter constraint \eqref{del1}, corresponding to the root $z_1=0$, are shown in Fig.~\ref{E1Delta} 
as functions of $\lambda$ and $\epsilon$ for specific values of $\Delta$ and $\omega$. 
In addition, the constraint \eqref{del1} as a function of $\Delta$ and $\epsilon$ is illustrated in Fig.~\ref{En1}.

\section*{Conclusions}

We have analyzed a generalization of the Rabi model, including two-photon and asymmetric terms, demonstrating that the problem can be viewed as a fourth-order ordinary differential equation in  Segal-Bragman space. Using the Bethe ansatz approach, we were able to find  analytical solutions to the problem, which appeared as a Fuchsian equation, for arbitrary states. 
As far as we know, the best way to extract information from this equation is through a quasi (conditionally) exact approach, bearing in mind that this approach entails restrictions between the model parameters, which is quite natural, since we are not dealing with an extremely complicated equation.

It is worth mentioning that we have a particular interest in applying the Bethe ansatz approach to multiphoton generalizations of the Rabi model, since, in the basic case, a Rabi model of $n$ photons yields an ordinary differential equation of order $2n$. We are currently working on extending this work to anisotropic, two-mode, and few-qubit cases.

\section*{Aknowledgment}
L.M.N. and S.Z. were supported by Spanish MCIN with funding from European Union Next Generation EU (PRTRC17.I1) and Consejeria de Educacion from JCyL through QCAYLE project, 
as well as grants PID2020-113406GB-I00 MTM funded by AEI/10.13039/501100011033, and RED2022-134301-T.

\appendix
\numberwithin{equation}{section}

\section{Appendix: Normalizability in Bargmann Space}\label{appendixA}

In Bargmann space, the inner product is defined as \cite{Bargman61}
\begin{equation}
\langle f, g \rangle = \frac1\pi \int_{\mathbb{C}} \overline{f(z)}\, g(z) \,e^{-|z|^2} d^2 z.
\end{equation}
We consider the function
\begin{equation}
\phi_1(z) = e^{\alpha z^2} \varphi(z),
\end{equation}
where \( \varphi(z) \) is a polynomial in $z$ and \( \alpha \in \mathbb{C} \).
To determine normalizability in Bargmann space, we require
\begin{equation}
 \frac1\pi  \int_{\mathbb{C}} \left| e^{\alpha z^2 } \varphi(z) \right|^2 \,e^{-|z|^2} d^2 z=  \frac1\pi  \int_{\mathbb{C}} \left|  \varphi(z) \right|^2 \, e^{2\text{Re}[\alpha z^2 ]-|z|^2} d^2 z < \infty.
\end{equation}
  As $ |z| \to \infty $, the polynomial $ \varphi(z) $ grows at most algebraically, hence, it suffices that the Gaussian weight $e^{2\text{Re}[\alpha z^2 ]-|z|^2}$ be integrable over $\mathbb{C}$. Using the basic relations $|z^2|=| z|^2$ and $\text{Re}[z]\leq |z| $, it is easy to show that $\text{Re}[\alpha z^2]\leq |\alpha z^2|\leq |\alpha || z|^2 $ for all $\alpha,\,z\in \mathbb{C}$. Therefore, the argument of the exponential fulfills the condition $2\text{Re}[\alpha z^2 ]-|z|^2  \leq (2|\alpha |-1)| z|^2$. Consequently, to ensure convergence of the integral, we require the coefficient of $| z|^2$ to be strictly negative, that is 
 \begin{equation}
|\alpha| < \frac{1}{2}.
\end{equation}

\end{document}